# Vibration Sensitivity of one-port and two-port MEMS microphones

Francis Doyon-D'Amour, Carly Stalder, Timothy Hodges, Michel Stephan, Lixiue Wu,
Triantafillos Koukoulas, Stephane Leahy, Raphael St-Gelais

**Abstract:** Micro-electro-mechanical system (MEMS) microphones (mics) with two acoustic ports are currently receiving considerable interest, with the promise of achieving higher directional sensitivity compared to traditional one-port architectures. However, measuring pressure differences in two-port microphones typically commands sensing elements that are softer than in one-port mics, and are therefore presumably more prone to interference from external vibration. Here we derive a universal expression for microphone sensitivity to vibration and we experimentally demonstrate its validity for several emerging two-port microphone technologies. We also perform vibration measurements on a one-port mic, thus providing a one-stop direct comparison between one-port and two-port sensing approaches. We find that the acoustically-referred vibration sensitivity of two-port MEMS mics, in units of measured acoustic pressure per external acceleration (i.e., Pascals per *g*), does not depend on the sensing element stiffness nor on its natural frequency. We also show that this vibration sensitivity in two-port mics is inversely proportional to frequency ($S_{\text{Pa}/g} \propto 1/f$) as opposed to the frequency independent behavior observed in one-port mics. This is confirmed experimentally for several types of microphone packages.

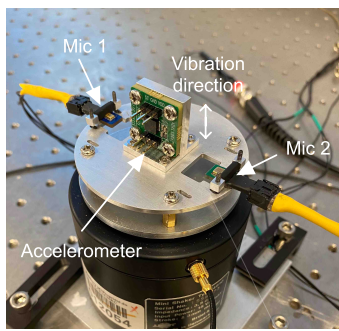
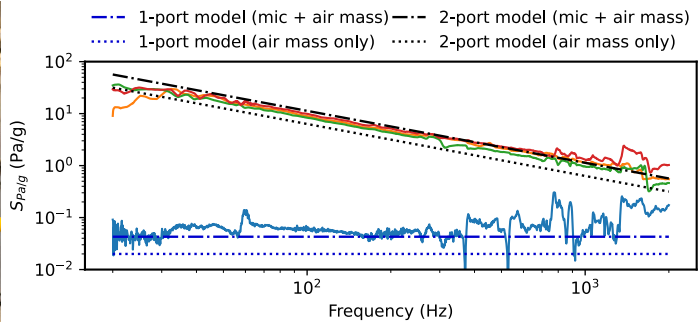

## I. INTRODUCTION

Two-port micro-electro-mechanical system (MEMS) microphones (mics) are currently receiving considerable interest, with the prospect of achieving higher directional sensitivity compared to traditional one-port microphones [1], [2]. Two-port MEMS mics employ a sensing element that is exposed to a sound wave from both sides—i.e., two acoustic ports, see Fig. 1 (a)—thus providing a measurement of the pressure difference between two locations in the sound wave ($\Delta P = P_2 - P_1$). This pressure difference strongly depends on the incoming angle of the sound wave and results in a figure-of-eight sensitivity pattern, maximized at 0° and 180°, that provides information on the sound wave direction (see Fig. 1b). Hybrid patterns between omnidirectional and figure-of-eight can also be obtained by adjusting the acoustic resistance of one of the ports [1], [3]. This directional sensitivity has potential applications in everyday electronics, augmented/virtual reality, online conferencing systems and more. In these applications, two-port MEMS mics can filter unwanted noises and achieve clearer audio recording in a smaller form factor [5].

While two-port mics provide information on the sound wave direction, they also detect much lower pressure signals than traditional one-port mics. In traditional one-port MEMS mics (schematized in Fig. 1 a), only one side of the flexible sensing element is exposed to the pressure of the sound wave ($P_1$) while the other is exposed to a reference pressure chamber ($P_{ref}$), typically at the average atmospheric pressure ($P_{atm}$). One-port MEMS mics therefore directly measure the acoustic sound pressure ($P_{acc} = P_1 - P_{atm}$). This quantity is fundamentally much larger than the pressure differential between two close points in the sound wave ($\Delta P = P_2 - P_1$), unless these two points are separated by at least one acoustic wavelength ($\lambda$). Such separation is typically not possible for MEMS assembly

F. Doyon-D'Amour, T. Hodges, M. Stephan and R. St-Gelais are with the Department of Mechanical Engineering, University of Ottawa, 75 Laurier Avenue East, Ottawa, Ontario K1N 6N5, Canada (email: fdoyo096@uottawa.ca; mstep054@uottawa.ca; thodg039@uottawa.ca; raphael.stgelais@uottawa.ca). T. Hodges, L. Wu and Triantafillos Koukoulas are with the National Research Council Canada, 1200 Montreal Road, Ottawa, Ontario K1A 0R6, Canada (email: Lixue.Wu@nrc-cnrc.gc.ca; Triantafillos.Koukoulas@nrc-cnrc.gc.ca). C. Stalder and S. Leahy are with Soundskrit, 1751 Rue Richardson #5102, Montreal, Quebec H3K 1G6, Canada (email: carly.stalder@soundskrit.ca; stephane.leahy@soundskrit.ca). R. St-Gelais is also affiliated with the Department of Physics, University of Ottawa, 75 Laurier Avenue East, Ottawa, Ontario K1N 6N5, Canada. The authors acknowledge financial support from Soundskrit and the Natural Science and Engineering Research Council of Canada (NSERC) through the Alliance Partnership Grant Program, Grant ALLRP 565214. F. Doyon-D'Amour also acknowledges support from the NSERC CREATE training program, Grant 497981.

in millimeter-scale packages, whereas the wavelength of sound is several centimeters (e.g., $\lambda \sim 34$ cm at $f = 1$ kHz). Measuring small pressure differences in two-port mics therefore requires the use of softer sensing elements that are presumably more prone to interference from external vibration. Nevertheless, vibration sensitivity of mics has only been previously characterized for one-port MEMS mics [7] – [11].

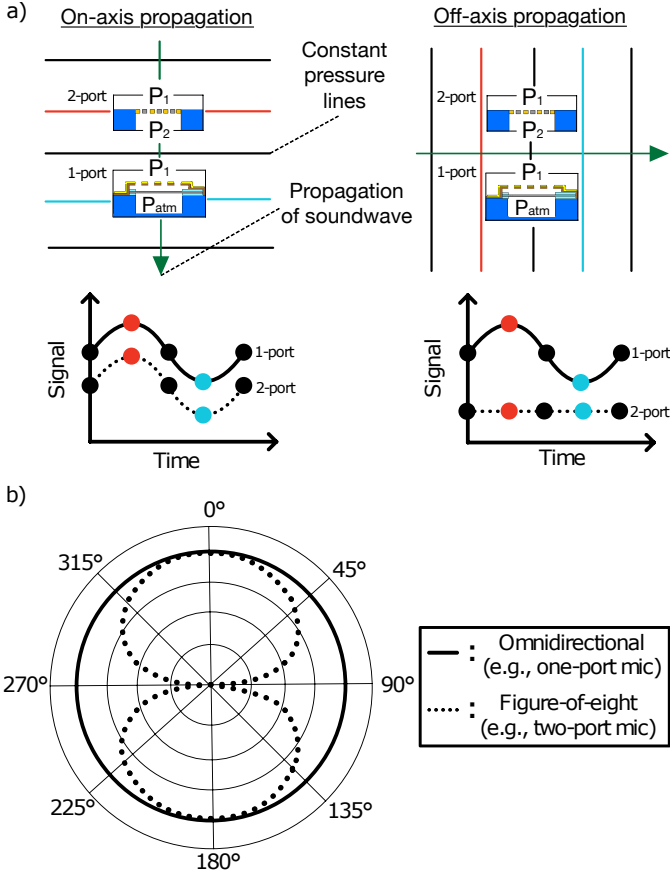

**Fig. 1.** (a) Schematic of sound waves impinging one-port and two-port mics at 0° (on-axis) and 90° incidence (off-axis). The signal picked up by the two-port mic is sensitive to orientation, as opposed to the one-port case. (b) Corresponding polar pick-up pattern for a one-port mic (omnidirectional) and a two-port mic (figure-of-eight).

Here we expand on two-port mics existing models for vibration sensitivity in one-port mics [10], and we experimentally demonstrate their validity for several emerging two-port microphone technologies. For completeness, we also perform experimental measurements on one-port MEMS mics, thus providing direct comparison between one-port and two-port MEMS mics in identical experimental conditions.

## II. THEORY

The vibration sensitivity of one-port MEMS mics can be accurately described by a lumped mechanical model [10] accounting for the mass and stiffness of the sensing element, together with the mass of air columns of length $L$, as shown in Fig. 2 (a). It is convenient to define this sensitivity in acoustically-referred terms, i.e., $S_{Pa/g}$ in units of measured acoustic pressure signal (in Pa) per external acceleration $g$ (where $g = 9.81$ m/s$^2$). $S_{Pa/g}$ thus provides the amount of parasitic acoustic signal (in Pa) recorded by the mic due to a given vibration signal (in $g$). A higher $S_{Pa/g}$ means that a mic is more sensitive to parasitic vibration, and as such lower $S_{Pa/g}$ is typically desired. When $L \ll \lambda$, $S_{Pa/g}$ is well approximated by [10]

$$S_{Pa/g,1port} = 9.81\rho_a(L_1 + 0.5L_2) + 9.81\rho_m t_m, \quad (1)$$

where $L_1$, $L_2$, are dimensions defined in Fig. 2 (a), $\rho_a$ is the density of air, $\rho_m$ and $t_m$ are the density and the thickness of the MEMS mic sensing element. In the common case where the mass of air has a more critical role than the mass of the microphone sensing element, (1) simplifies further to

$$S_{Pa/g,1port} \approx 9.81\rho_a(L_1 + 0.5L_2). \quad (2)$$

Using the same approach, we define a lumped mechanical model for the case of a two-port mic. We consider the dimensions illustrated in Fig. 2 (b), and we assume a lumped mass that accounts for the mass of the mic sensing element and the air present on both sides:

$$m_{tot,2port} = \rho_a A(L_1 + L_2) + \rho_m t_m A, \quad (3)$$

where $A$ is the area of the sensing element.

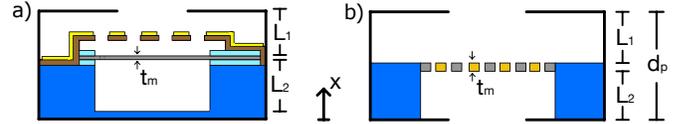

**Fig. 2.** Cross-section view of (a) a one-port mic and (b) a two-port mic showing distance ($L_1$, $L_2$) between the sensing element and the ports. Other geometries for two-port mics are given in appendix section A.

We then express the pressure difference between the two ports of a directional mic ($\Delta P = P_1 - P_2$) as a function of the absolute acoustic pressure ($P_{acc} = P_1 - P_{atm}$). We consider sound propagation along the $x$-direction, and on-axis with the two ports (e.g., left panel of Fig. 1 a). The sound wave pressure at a specific point in space is described by the wave equation

$$P = P_{acc} e^{i(kx-\omega t)} + P_{atm}, \quad (4)$$

where $P$ is the pressure at location $x$ and at time $t$, $P_{acc}$ is the acoustic pressure, $\omega$ is the frequency and $k$ is the wave number of the sound wave ($k = \omega/c$, where $c$ is the speed of sound). To estimate the pressure difference between two close points, we differentiate (4) with respect to position $x$

$$\frac{\partial P}{\partial x} = \frac{\omega}{c} P_{acc} e^{i(kx-\omega t)}. \quad (5)$$

In cases where the distance between the ports is smaller than the wavelength of the incident sound wave ($d_p \ll \lambda$), we can assume a linear change of pressure between the ports using $\partial x \approx d_p$ and $\Delta P \approx \partial P$. This yields

$$\Delta P = \frac{P_{acc} d_p \omega}{c}. \quad (6)$$

We then study separately the displacement sensitivity to vibration ($S_{m/g}$), and the displacement sensitivity to acoustic pressure ($S_{m/Pa}$). These two sensitivities can later be combined in an acoustic-referred vibration sensitivity ($S_{Pa/g}$) for a two-port mic

$$S_{\text{Pa}/g,2port} = \frac{S_{m/g,2port}}{S_{m/\text{Pa},2port}}. \quad (7)$$

For compactness, we do not carry the "2port" notation from now on (e.g., $S_{\text{Pa}/g} \equiv S_{\text{Pa}/g,2port}$). The following therefore only applies to two-port mics.

The acoustic displacement amplitude ($X_{acc}$) of the mic sensing element subjected to the pressure signal $\Delta P$ is given by:

$$X_{acc}(\omega) = \chi(\omega)\, \Delta P \cdot A, \quad (8)$$

where $\chi$ is the dynamic mechanical response (in m/N) of a harmonic oscillator of natural frequency $\omega_n$, quality factor Q and mass of sensing element $m$ [12]:
$\chi(\omega) = \left( m\omega_n^2 \sqrt{\left(1 - \frac{\omega^2}{\omega_n^2}\right)^2 + \left(\frac{\omega}{Q\omega_n}\right)^2} \right)^{-1}$. Using (6) and (8), the displacement sensitivity to acoustic pressure ($S_{m/\text{Pa}}$) is then given by:

$$S_{m/\text{Pa}}(\omega) \equiv \frac{X_{acc}(\omega)}{P_{acc}} = \frac{d_p \omega A}{c} \chi(\omega). \quad (9)$$

Likewise, the sensing element displacement ($X_g$) due to external acceleration $a$ can be described by a mass-spring system subjected to an inertial force ($F_g$). This yield

$$X_g(\omega) = \chi(\omega)\, F_g = \chi(\omega)\, m_{tot} a, \quad (10)$$

where $m_{tot}$ is given in (3). After rearranging (10), the displacement sensitivity to vibration ($S_{m/g}$) becomes:

$$S_{m/g}(\omega) = \frac{X_g(\omega)}{a} = 9.81\, m_{tot}\, \chi(\omega), \quad (11)$$

We can finally combine $S_{m/\text{Pa}}$ from (9) and $S_{m/g}$ from (11) to obtain an expression for the acoustic-referred vibration sensitivity (in Pa/g) of a two-port mic:

$$S_{\text{Pa}/g}(\omega) = \frac{S_{m/g}}{S_{m/\text{Pa}}} = \frac{9.81\,(\rho_m t_m + \rho_a(L_1 + L_2))\, c}{d_p \omega}. \quad (12)$$

When the mass of air is more important than the mass of the mic sensing element, $\rho_a(L_1 + L_2)) \gg \rho_m t_m$, and when $d_p = L_1 + L_2$ (e.g., see Fig. 2 b), (12) simplifies further to:

$$S_{\text{Pa}/g}(\omega) \approx \frac{9.81\, \rho_a\, c}{\omega}. \quad (13)$$

Cases in which $d_p \neq L_1 + L_2$ are discussed further in the results and in Appendix section A.

Interestingly, we find in (12), (13) that two-port mics can be expected to have an acoustic-referred sensitivity to vibration that is inversely proportional to frequency ($S_{\text{Pa}/g} \propto 1/\omega$) as opposed to a frequency independent behavior of one-port mics observed in (1), (2). We also find that while the mic sensing element stiffness influences the raw sensitivity to vibration, through $\chi$ in (11), it cancels out in the final acoustic-referred vibration sensitivity in (12).

In the context of experimentally characterizing the vibration sensitivity, the link between the sensing element displacement ($X_{acc}$ and $X_g$ in m) and the mic raw signal (in V) is not immediately available to most microphone end users. In this case we recover the acoustic-referred vibration sensitivity ($S_{\text{Pa}/g}$) by normalizing the raw measured acceleration response ($S_{V/g}$) by the acoustic sensitivity that is either measured or given by the microphone manufacturer ($S_{V/\text{Pa}}$):

$$S_{\text{Pa}/g}(\omega) = \frac{S_{V/g}}{S_{V/\text{Pa}}} \quad (14)$$

### III. EXPERIMENTAL APPARATUS AND METHOD

We validate the model of section II using a custom mechanical shaker apparatus on which we can measure the vibration response of various one-port and two-port mics. Vibration is generated using an electrodynamic exciter (B&K Type 4810 mini-shaker) controlled by an arbitrary function generator. The signal of the function generator (typically < 400 mV) is pre-amplified by 20 dB with a power amplifier (B&K Type 2706). The acceleration of the electrodynamic exciter is measured by a one-axis analog accelerometer (Analog Device ADXL 1005) mounted on a custom-made aluminum plate (see Fig. 3) on which up to two mics under test can be attached. The plate comprises openings to prevent blocking of the microphone ports, as well as mounting hardware for measuring either on-axis or off-axis vibration sensitivity (see Fig. 3 b). On-axis measurements refer to vibrations that are parallel to the ports spacing of the mics, while off-axis measurements refer to vibrations that are perpendicular to the ports spacing. All mics characterized in this work are mounted on custom-made printed circuit boards that are designed with a compact form factor for ease of mounting and testing (e.g., see Fig. 3 b). Additional details on the microphone packages are given in Appendix section A.

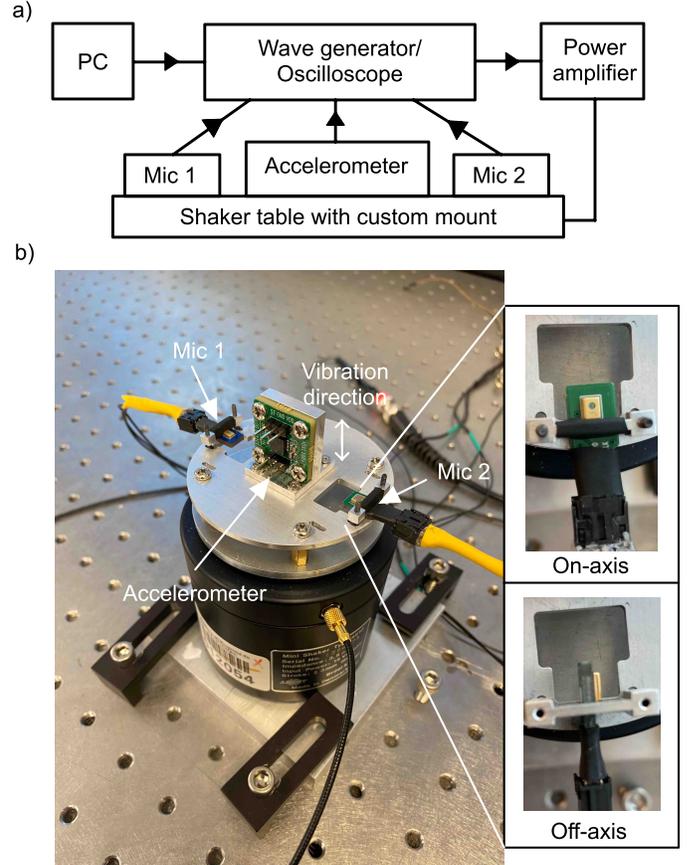

**Fig. 3.** (a) Bloc diagram of the experimental setup (b) Picture of the electrodynamic exciter with a custom mount for an accelerometer and two MEMS microphones under test. The mics are either mounted on-axis or off-axis, as shown in the insets (right).

We characterize the acceleration response of the electrodynamic shaker by sweeping the excitation frequency from 20Hz to 20kHz (i.e., the audible frequency range), while keeping the excitation signal amplitude constant. The corresponding shaker acceleration response, measured by the reference accelerometer, is shown in Fig. 4 (a). These responses are roughly constant in the 100 – 1000 Hz range. Deviations from a flat response at low frequencies (<100 Hz) are caused by the limited shaker amplitude [13]. At higher frequencies (>1 kHz) deviation from a flat response are caused by the first mechanical resonance mode of our mounting plate. This resonance occurs at ~2.5 kHz, as expected theoretically (see Appendix section B) [14]. These deviations from an ideally flat frequency response are found to be highly repeatable and are compensated-for by the accelerometer measurement in all our subsequent tests. As extra caution, we also limit measurement in the 20 – 2000 Hz frequency range. Note that this low end of the acoustic frequency range is where the mic sensitivity to vibration is expected to have the most practical importance, as shown in (12), (13). The shaker acceleration response is found to be highly predictable, repeatable and linear as a function of the excitation potential. This is well illustrated in Fig. 4 (b), where the measured acceleration increases linearly with the excitation amplitude, regardless of the excitation frequency.

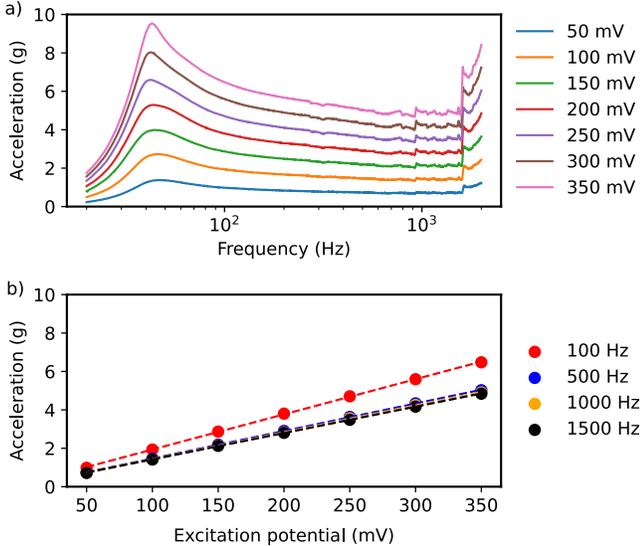

**Fig. 4.** Electrodynamic exciter characterization. (a) Acceleration measured by the accelerometer on the electrodynamic exciter for different signal excitation amplitudes. The roll off at frequencies < 40 Hz is caused by the limited shaker travel range. (b) Linearity of acceleration vs signal amplitude, at different frequencies.

## IV. RESULTS

In this section, the acoustically-referred vibration sensitivities of various mics are experimentally characterized and compared with theoretical models. Table 1 summarizes the results from all the mics under test and their estimated relevant parameters needed for the theoretical predictions. The air column lengths ($L_1, L_2$) and the port spacing ($d_p$) are measured from the mics packages and can vary between models. The Soundskrit Inc. SKR0400 and TDK Invensense Inc. ICS40800 mics are well illustrated by the configuration of Fig. 2 (b) (see also Appendix Fig. S1), while the Knowles Electronics, LLC. Lazarus mic is well represented by Fig. 2 (a) (see also Appendix Fig. S2). We also include mics with different spatial configurations to test the limits of validity of our model. A HARMAN International Hendrix 120 Topaz mic comprise a V-shaped package which essentially changes the effective port spacing ($d_p$) and air column lengths ($L_1, L_2$), as shown in Appendix Fig. S3. Likewise, we also test a custom-built array of two Knowles Electronics, LLC. Winfrey omnidirectional mics wired to a differential amplifier to create an effective two-port mic, shown in Appendix Fig. S5. Note that for compactness, the legal suffixes (e.g., "Inc"), the part number, and the full legal name of microphone manufacturers are omitted after being mentioned in this paragraph and in Table 1.

We measure the raw sensitivity to acceleration ($S_{V/g}$), which can later be used to obtain $S_{Pa/g}$ using (14). Fig. 5 (a) shows the raw measured sensitivity to vibration ($S_{V/g}$) of the five mics under test. We note that the mics responses ($S_{V/g}$) are relatively flat since our measurement frequency upper limit is smaller than the mics resonance frequencies (from ~4.5 kHz for Soundskrit, to ~39 $kHz$ for Lazarus). As expected, the raw sensitivity to vibration scales as ~$1/\omega_n^2$, and is therefore maximum for the Soundskrit mic.

Table 1: Mics tested and their estimated parameters.

| Microphones | $L_1$ (mm) | $L_2$ (mm) | $d_p$ (mm) | $\omega_n$ (kHz) |
|---|---|---|---|---|
| One-port mic (1p) | | | | |
| Knowles Lazarus [15] (SPH18C3LM4H-1) | 1.25 | 1.25 | N/A | 39 |
| Two-port mic (2p) | | | | |
| Soundskrit [5] (SKR0400) | 1.25 | 1.25 | 2.5 | 4.5 |
| TDK [4] (ICS48000) | 1.25 | 1.25 | 2.5 | 15 |
| Harman Hendrix 120 Topaz [16] (E678-0002-003) | 7.5 | 7.5 | 12 | 10 |
| Array of two one-port mics | | | | |
| Knowles Winfrey [17] (SPM0687LR5H-1) | 1.25 | 1.25 | 10 | 25 |

We then plot, in Fig. 5 (b) the acoustic sensitivities ($S_{V/Pa}$) of the different mics. While these sensitivities can typically be obtained from the manufacturer specification sheets [4], [5], [15] – [17], we opt for characterizing each of them individually using a dedicated acoustic testing apparatus. We therefore account for hypothetical deviations from the mics specifications. The measured acoustic sensitivities are within 0.3 dB (at 1 kHz) of the manufacturer specifications. More details on the microphone interfacing and testing conditions are given in Appendix section A (Microphones Packages) and

Appendix section C (Test setup for microphone sensitivity calibration).

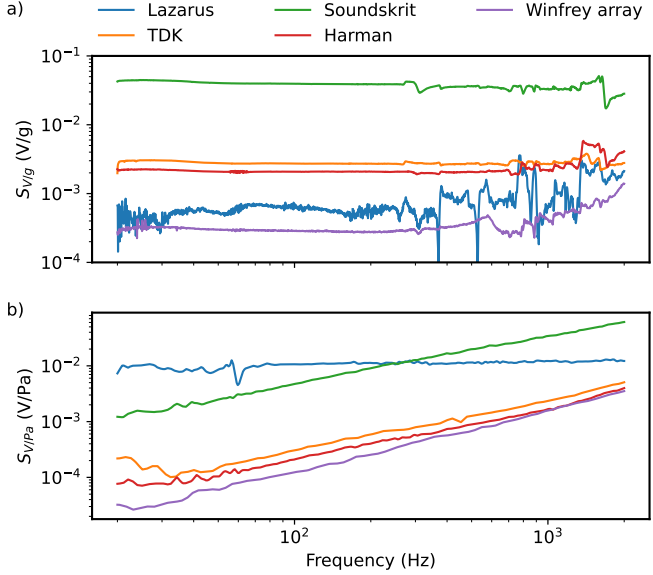

**Fig. 5.** (a) Raw measured vibration sensitivities in V/g ($S_{V/g}$), and (b) Measured acoustic sensitivities in V/Pa ($S_{V/Pa}$) for all mics under test.

By combining the measured acoustic sensitivity (Fig. 5 b) and raw measured vibration sensitivity (Fig. 5 a) using (14) we finally obtain, in Fig. 6 (a), our final metric of interest: the acoustically-referred vibration sensitivity ($S_{Pa/g}$). As expected from our model, the acoustically-referred vibration sensitivity ($S_{Pa/g}$) of two-port mics is inversely proportional to frequency in all cases, while it is frequency independent for the one-port mic. Fig. 6 (a) also provides a direct comparison of these results with our theoretical model shown in (1), (2), (12), (13). We obtain the best correspondence with our model by assuming a membrane thickness ($t_m \sim 1$ µm) and density ($\rho_m \sim 2300$ kg/m$^3$, i.e.– silicon) typical of common MEMS mics [18]. In this case, we find that the one-port and two-port mics in the configuration of Fig. 2 (i.e., the Knowles Lazarus one-port, and the Soundskrit and TDK two-port) match almost exactly with our developed models. This is also the case for the Harman mic, even though the ports do not strictly match the sum of the air column lengths in this case, i.e., $d_p \neq L_1 + L_2$ (see microphone package schematic in Fig. S3, and dimensions in Table 1). The mismatch is likely to be too small to make a noticeable difference in this case.

Small differences between the Soundskrit, TDK and Harman mic vibration sensitivity responses can be attributed to our estimated dimensions for the microphone package as well as their estimated membrane thicknesses and density. Upper and lower uncertainty bounds for our model are presented in appendix section D and are found to match well with our results.

The only noticeable deviation from our models is observed for the custom-made array of two Knowles Winfrey mics, most likely due to the use of a dual sensing element that make the air column lengths ($L_1, L_2$) dramatically different than the acoustic port spacing $d_p$ (see also Fig. S5). Nevertheless $S_{Pa/g}$ still rigorously follows a $1/f$ behavior (see Fig. 6 a) in this case. We can also define an effective air column length (on the order of $L_{eff} \cong \frac{d_p}{5}$) for which (12) matches our measurement exactly. Explanation of this factor 5 would likely be possible with further hydrodynamic modeling, but is beyond the scope of the current work.

The responses of the mics when subjected to off-axis vibrations (i.e., as in Fig. 3 b whereas the vibration direction is in line with the least acoustically sensitive orientations of the two-port mics) are also measured and shown in Fig. 6 (b). In this case, we find that $S_{Pa/g}$ is significantly lower, which is expected given that the air columns are no longer aligned with the vibration direction. On the other hand, for one-port mics, vibration direction essentially does not affect $S_{Pa/g}$. This may indicate that one-port mics pick up parasitic sound from the electrodynamic exciter in addition to the inertial vibration signal. This hypothesis is explored in appendix section E, where we find that the measured signal of two-port mics is clearly dominated by the vibration signal, whereas the effect of sound from the exciter cannot be entirely ruled out for one-port mic in the current experiment. Readers interested in one-port mics vibration sensitivity should likely consult work dedicated specifically to one-port mics characterization [7] – [11], even though we note that the one-port measurements presented herein agree with the ones reported in the literature [10].

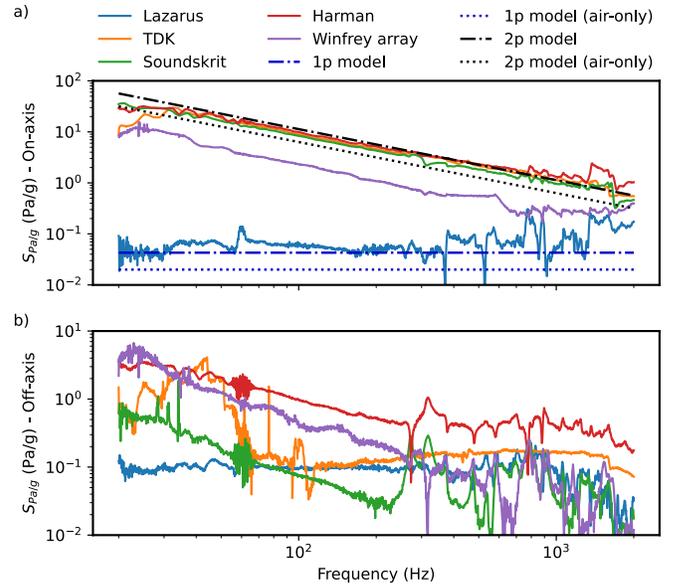

**Fig. 6.** Measured acoustically-referred vibration sensitivity in Pa/g ($S_{Pa/g}$), and comparison with the developed models. Air-only models account for the mass of the air column only (13), neglecting the mass of the sensing element (12). Panel (a) present measurements where the vibration direction is in line with the mics port spacing (i.e., "on-axis"). Panel (b) present measurements of vibrations in the perpendicular direction (i.e., "off-axis").

## V. Conclusion

The presented models accurately predict the vibration sensitivity of one-port and two-port MEMS microphones. We demonstrate that the acoustically-referred vibration sensitivity of two-port mics is inversely proportional to frequency ($S_{Pa/g} \propto 1/f$), whereas it is frequency independent for one-port mics. This behavior was observed regardless of the tested microphone acoustic package types. Interestingly, the acoustic-referred vibration sensitivity is found to be independent of the microphone natural frequency (i.e., of its stiffness). Microphones with softer sensing elements are more sensitive to vibration, but also equally more sensitive to sound, thus making the acoustically-referred vibration sensitivity constant. When working with two-port microphones, acoustic designers should therefore expect similar spurious acoustic signals resulting from vibrations, regardless of the mic model and manufacturer.

## APPENDIX

### A. Microphones packages

This section provides details on all the microphones used in this work, with a schematic showing the dimensions used in Table 1 in the main text (i.e., the dimensions used for our theoretical model).

#### 1) Soundskrit, TDK

The Soundskrit and TDK microphone packages designs are shown in Fig. S1. The package geometry is the closest to the baseline two-port package shown in Fig. 2 in the main text. They have a sensing element open to the sound wave pressure on both sides.

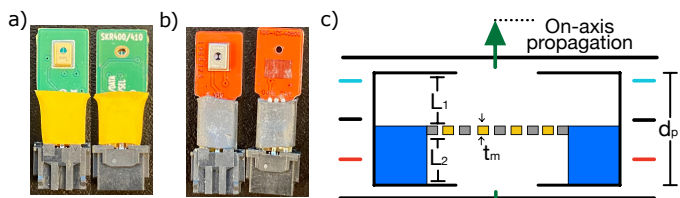

**Fig. S1.** (a) Soundskrit SKR0400 microphone (b) TDK ICS40800 microphone (c) generalized dimensions for general two-port microphones.

#### 2) Lazarus

The Knowles Lazarus microphone package (Fig. S2) is used for our main one-port theoretical model. The sensing element is only open to the sound wave pressure on one side.

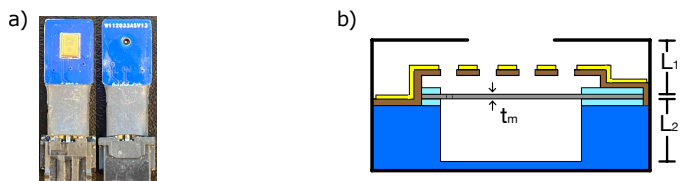

**Fig. S2.** (a) Knowles Lazarus microphone (b) generalized dimensions for general one-port microphones.

#### 3) Harman Hendrix 120 Topaz

The Harman Hendrix 120 Topaz has a sensing element open to the sound wave pressure with a V shaped package which, as shown in Fig. S3, changes effective port spacing and air column lengths. The original, as-specified mic [16], had a partial acoustic resistance (i.e., a mesh) converting the figure-of-eight polar pattern into a cardioid, in a fashion similar to [1], [3]. For this work, this meshing was removed, such that all characterized two-port microphones have figure-of-eight directional sensitivity.

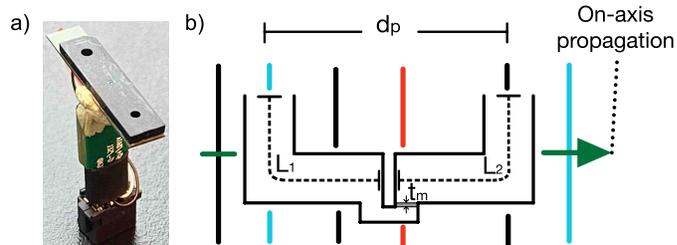

**Fig. S3.** (a) Harman Hendrix 120 Topaz microphone (b) generalized dimensions of Harman Hendrix 120 Topaz

Removing the mesh did not affect the raw voltage responses generated by vibration (see Fig S4 a). However, as expected, the acoustic sensitivity ($S_{V/Pa}$, see Fig. S4 b) decreased after mesh removal, leading to different acoustic-referred vibration sensitivity ($S_{Pa/g}$).

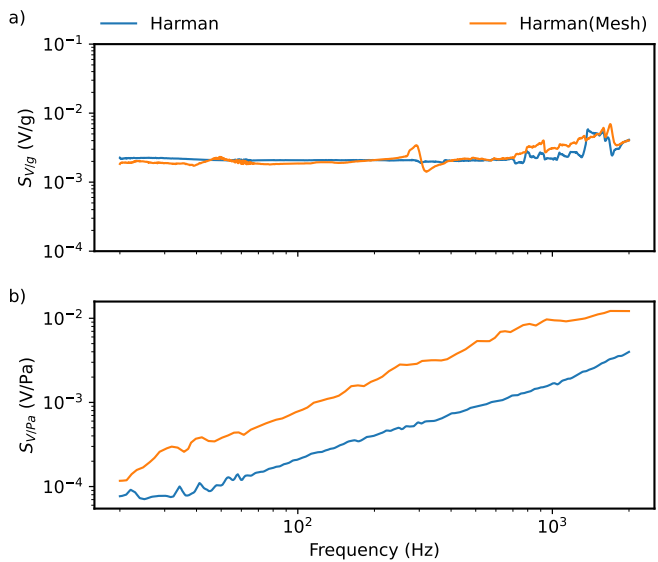

**Fig. S4.** Comparison of raw voltage and response and sensitivity of the Harman mic with and without an acoustic resistance (mesh)

#### 4) Array of two Knowles Winfrey

Fig. S5 presents the custom-built array of two Knowles Winfrey omnidirectional microphones mounted 10 mm away on a custom PCB to create an effective two-port mic. Both mics have one-port open to the pressure of the sound wave and are then differentially wired for measuring pressure gradients and creating an effective two-port mic.

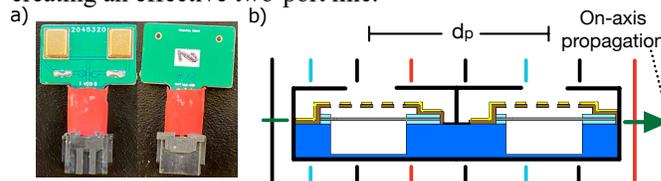

**Fig. S5.** (a) Array of two Knowles Winfrey microphone (b) generalized dimensions of the array.

### B. Mounting plate resonance

We estimate our mount's natural frequency by approximating it as a circular plate simply supported around its circumference. The natural frequency of the fundamental mode is then given by [14]:

$$\omega_n = \frac{4.979}{r^2}\sqrt{\frac{D}{\rho h}}, \quad (S1)$$

where $r$ is the radius of the plate, $D$ is the plate stiffness, $\rho$ is the density of the plate and $h$ is the thickness of the plate. To calculate the plate stiffness, we use [14]:

$$D = \frac{Eh^3}{12(1-\mu^2)}, \quad (S2)$$

where $E$ is the Young's modulus, and $\mu$ is the poisson ratio. We calculated an estimated plate resonance of 2.5 kHz which is close to what we observe experimentally (see Fig. S6).

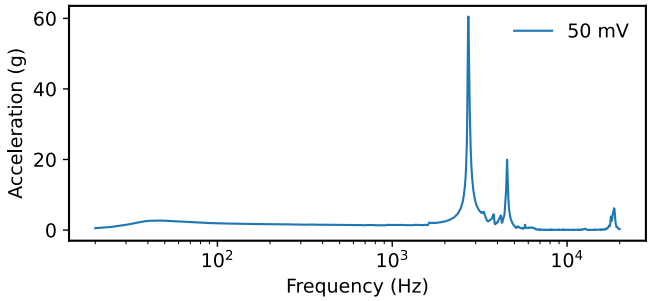

**Fig. S6.** Mounting plate frequency sweep shows a resonance around 2.5 kHz.

## C. Test setup for microphone sensitivity calibration

A setup at Soundskrit headquarters allows acoustic sensitivity measurements to account for potential deviations from datasheets. It consists of two reference microphones, an audio interface, an anechoic chamber, and a speaker. The speaker sweeps tones from 20 Hz to 20 kHz for a reference mic, and again for the microphone under test. Table S1 shows the measured sensitivity at 1 kHz, compared with the specified datasheet sensitivity, both at 1 kHz. Special notes are applicable for some of these tests. In the TDK (ICS40800) mic datasheet, the acoustic sensitivity is measured with one of the acoustic ports blocked (i.e., essentially creating an omnidirectional one-port mic) [3]. Hence, a measurement with one acoustic port blocked was performed to confirm accordance with the -38 dBV/Pa sensitivity from the datasheet. A second measurement, in the two-port configuration, yielded -52.6 dBV/Pa and is used in current work. For the Harman Hendrix, an acoustic resistance (mesh) was removed for our experiment. Sensitivity measurements were performed before mesh removal to confirm specified performance (-40.5 dBV/Pa) before, and also after mesh removal (-55.5 dBV/Pa). There is no datasheet sensitivity for the array of two Knowles Winfrey mics since it is a custom-made device.

Table S1: Measured versus datasheet sensitivity.

| Mics | Soundskrit | TDK | Harman |
|---|---|---|---|
| Datasheet sensitivity at 1 kHz (dBV/Pa) | -29 | -38[1] | -40.5[2] |
| Measured sensitivity at 1 kHz (dBV/Pa) | -29.3 | -37.8[1] (-52.6)[3] | -40.5[2] (-55.5)[3] |

| Mics | Lazarus | Array of two Knowles Winfrey |
|---|---|---|
| Datasheet sensitivity at 1 kHz (dBV/Pa) | -38 | N/A |
| Measured sensitivity at 1 kHz (dBV/Pa) | -38.3 | -34.5 |

[1] Measured with one acoustic port sealed
[2] Measured with partial acoustic resistance
[3] As used in this work (two opened acoustic ports)

## D. Uncertainty estimations

Our models rely on various estimated parameters that contribute to uncertainty, such as the air column lengths, the membrane thickness, and its density. In Table S2 we provide estimated confidence intervals for these values, which we then used for uncertainty estimations. From these values, we plot the upper and lower limits of our model alongside our experimental results. The upper limit is 62% higher than our original model, whereas the lower limit is 28% lower, as shown in Fig. S7. Experimental data essentially fall within this confidence interval except at the extreme frequency limits of measurements. In these cases, deviation most likely results from challenges in measuring acoustic sensitivity at low frequencies, and from the presence of the mounting plate resonances at high frequencies.

Table S2: Mic parameters confidence intervals.

| Air columns lengths ($L_1, L_2$) [mm] | Membrane density ($\rho_m$) [kg/m$^3$] | Membrane thickness ($t_m$) [μm] |
|---|---|---|
| 1 – 1.5 | 2000 – 3000 | 0.5 – 1.5 |

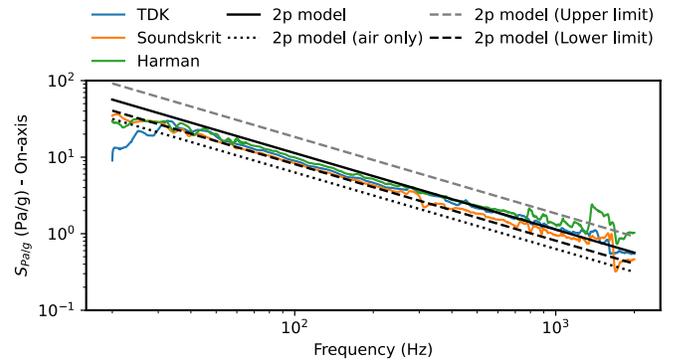

**Fig. S7.** Confidence interval for our theoretical model, showing good agreement with experimental results.

## E. Off-shaker measurements

To understand the potential contribution of sound emitted by the electrodynamic exciter, additional mics are mounted off the shaker (i.e., not subjected to vibrations) at about 1 cm above the mounting plate. Their responses are compared with the mics mounted on the shaker (i.e., subjected to vibrations). If the shaker sound (rather than vibration) is significant, the ratio between the mounted-on and mounted-off response would be close to 1. The ratio from typical two-port mics is clearly dominated by vibration with mounted-on mics responses being 4 to 6 times higher than mounted-off mics. However, the array of two Knowles Winfrey mics, and the Lazarus omnidirectional mics are close to a value of 1 (see Fig. S8). This may indicate that the actual sensitivity to vibration of one-port mic measured in this work is overestimated, even though it is in agreement with previous work [10].

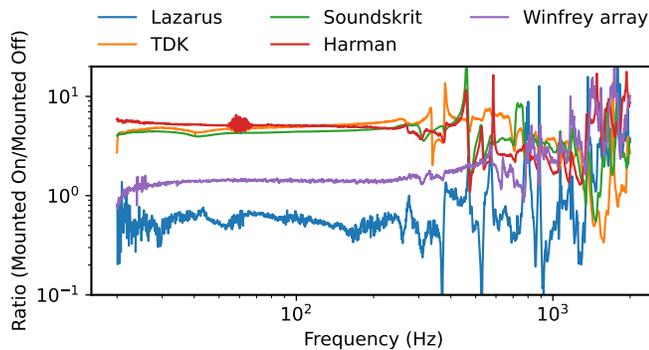

**Fig. S8.** Ratio of microphone responses while mounted on and off the shaker.